\documentclass[fleqn,twoside,twocolumn,nofootinbib,showkeys,11pt]{revtex4} 
\usepackage{verbatim}

\usepackage{cleveref}%
\usepackage{cmap} 
\usepackage[cp1251]{inputenc}
\usepackage[english]{babel}
\usepackage[T2A]{fontenc}
\usepackage{amsmath}
\usepackage{amstext}
\usepackage{amssymb}
\usepackage{bm}
\usepackage[pdftex]{color,graphicx}%
\usepackage[pdftex,plainpages=false]{hyperref}%

\begin{document}
\title[Magnetic Isotope Effect in the Uranium Isotopes Separation]
{MAGNETIC ISOTOPE EFFECT IN THE URANIUM ISOTOPES SEPARATION}%

\author{I.\,V.\,Zhyganiuk}
\affiliation{Institute for Safety Problems of Nuclear
Power Plants,   Nat. Acad. of Sci. of Ukraine}
\address{36a, Kirova str., Chornobyl 07270, Ukraine}
\email{i.zhyganiuk@ispnpp.kiev.ua}
\affiliation{ Institute of Artificial Intelligence Problems,
Nat. Acad. of Sci. of Ukraine}
\address{40, Academician Glushkov Ave., Kyiv 03680, Ukraine}
\author{A.\,V.\,Zubko}%
\affiliation{Institute of Environmental Geochemistry, Nat. Acad. of Sci. of Ukraine}
\address{34a, Palladin Ave., Kyiv 03142, Ukraine}
\author{O.\,B.\,Lysenko}%
\affiliation{Institute of Environmental Geochemistry, Nat. Acad. of Sci. of Ukraine}
\address{34a, Palladin Ave., Kyiv 03142, Ukraine}




\begin{abstract}
When uranyl nitrate is photolysed in the water solution, light isotope  ${}^{235}{\rm U}$ is separated from isotope ${}^{238}{\rm U}$ ; an enrichment factor $K = 1.04$ . The initial samples were depleted by a light isotope of uranium ${}^{235}{\rm U}$  (degree $\beta  = 0.00455$). Regenerated uranyl nitrate was enriched by an isotope of uranium ${}^{235}{\rm U}$  (degree $\alpha  = 0.00464$). Uranium tetrafluoride was depleted by an isotope of uranium ${}^{235}{\rm U}$   (degree $\beta  = 0.00446$). Uranium isotopes are separated by the magnetic isotope effect (MIE) in uranyl nitrate photoreduction.
\end{abstract}


\maketitle
\thispagestyle{empty}

\begin{flushright}
{\footnotesize\it In memory of Academician Emlen Sobotovich, an outstanding scientist\\
}
\end{flushright}

\section{Introduction}

The results of a theoretical investigation of interactions between electron magnetic moments and
nuclear--magnetic moment published in work~\citep{Evans}
 in 1975; this work was the first to show that existence to the MIE.
Buchachenko\,A. and his group originally discovered the MIE in 1976~\citep{2Buchachenko}.

Uranyl nitrate ions complexes  are being regenerated uranyl-ion and nitrate-ion radicals   pairs    induced by photolysis in heavy water solutions; authors investigated MIE in these chemical reactions.  These results are presented in this manuscript. For more details about MIE, see~\citep{2Buchachenko,Sobotovich,4Buchachenko,5Buchachenko,Lysenko,Skulskii,Korkushko,6Buchachenko,Sulaberidze}. These effects and processes have been used for a new chemical separation of uranium isotopes.

\section{Experimental}

In the photoinduced reaction uranyl nitrate was regenerated in the р-methoxyphenol and heavy water solution. The initial samples of uranyl nitrate were depleted by a light isotope of uranium ${}^{235}{\rm U}$ (degree $\beta  = 0.00455$). In work~\citep{6Buchachenko}, the initial samples of uranyl nitrate were enriched by an isotope of uranium ${}^{235}{\rm U}$ (degree $\alpha  = 0.112$). Our experiments were conducted in solutions of following concentration: ${\rm UO}_2 ({\rm NO}_3 )_2$ --- ${\rm 1}{\rm .6} \cdot {\rm 10}^{{\rm  - 3}} $ mol/l, [р-methoxyphenol] --- ${\rm 5}{\rm .0} \cdot {\rm 10}^{{\rm  - 2}}$, ${\rm NH}_4 {\rm F}$ --- ${\rm 1}{\rm .0}$ mol/l, ${\rm H}_2 {\rm SO}_4$ --- ${\rm 0}{\rm .5}$ mol/l in heavy water.

Depleted uranium dioxide was dissolved in concentrated nitric acid; this solution was evaporated; we had received uranyl nitrate in dry residue. Solutions were prepared and mixed; they were placed in a quartz vessel. Next step, the solutions were deoxygenated flushing with argon; the quartz vessel was hermetically sealed.

Further, the solutions had been irradiating by an ultraviolet lamp (characteristics of the lamp: $\Delta \lambda  = 350 \div 450 \; nm,  P=61 W $), until a suspension with  ${\rm UF}_4$ was  formed in solutions. Next, the irradiation had been stopping, before the precipitate with ${\rm UF}_4$    didn’t completed formation. Then, the suspension with ${\rm UF}_4 $   was being decanted in the centrifuge. Here, we had separated uranium tetrafluoride in the precipitate. Uranium isotope ratios were being measured in the uranium tetrafluoride and the uranyl nitrate. These results were being obtained by MI-1201 mass spectrometer (at State Institution ''Institute of Environmental Geochemistry, National Academy of Sciences of Ukraine'').

\section{Results and their discussion}

\begin{figure}
\vskip1mm
\includegraphics[height=5cm]{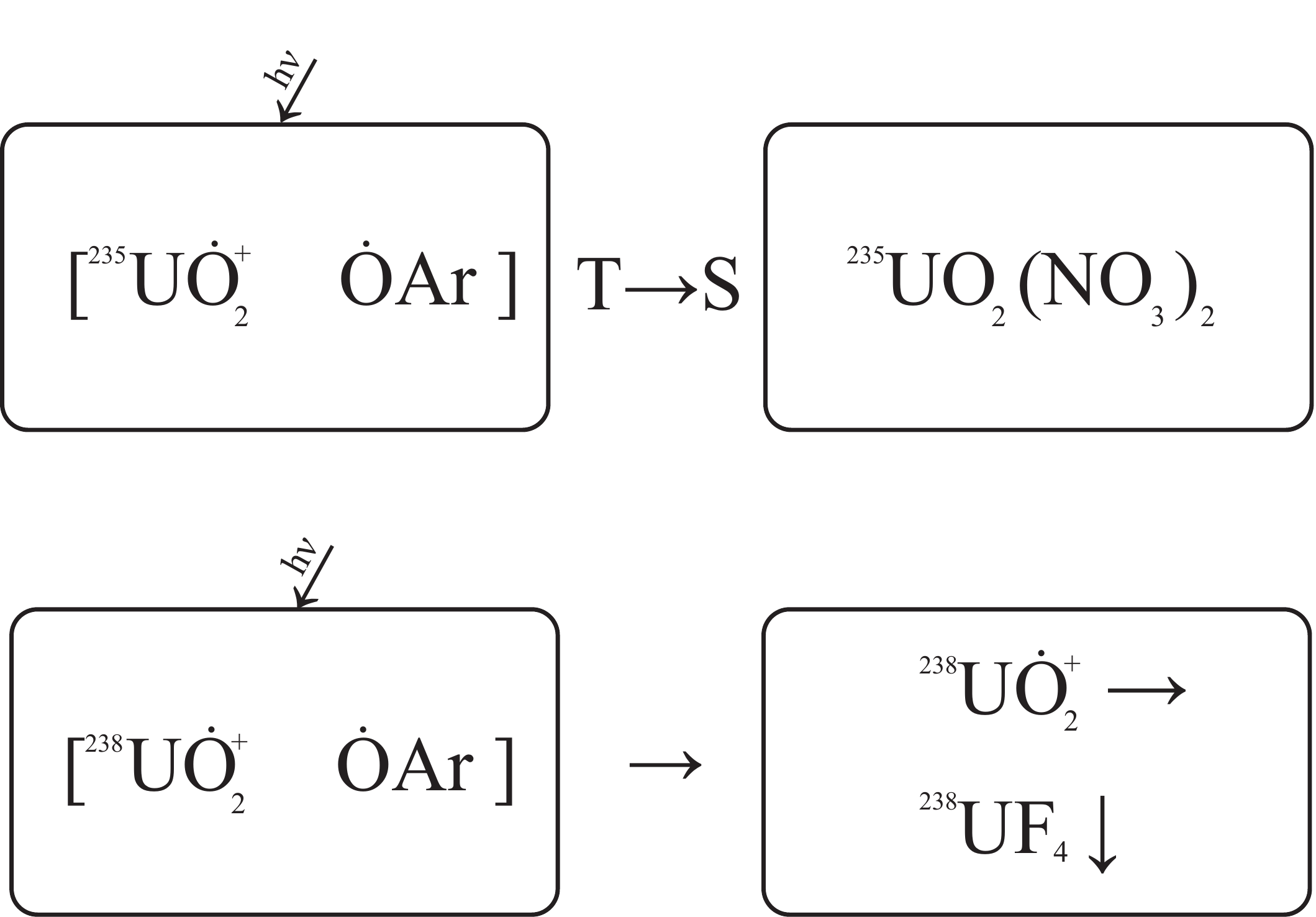}
\vskip-1mm
\caption{Scheme of the processes were being occurred in the photolysis of uranyl; in this scheme introduce next denotation: unpaired electron $(\cdot)$, replaced phenol $(\rm {Ar})$, triplet state $(\rm {T})$, singlet state $(\rm {S})$, uranoyl $({\rm UO}_2^ +)$}\label{f1}
\end{figure}

A triplet-singlet conversion had been finished in radical pair before the uranyl regenerated to the uranyl nitrate; it is presented by Fig.~\ref{f1}. The triplet-singlet conversion has been occurred because electrons spines precessed in magnetic field of uranium nucleus ${}^{235}{\rm U}$. An uranoyl  ${\rm UO}_2^ +$ is a singly charged complex of uranyl-ion with unpaired electron. At this stage of reaction, an uranyl-ion ${\rm UO}_2^ {2+}$ is being enriched by isotope ${}^{235}{\rm U}$; the uranoyl ${\rm UO}_2^ +$ is being depleted by isotope ${}^{235}{\rm U}$. Uranoyl formed the insoluble precipitate with uranium tetrafluoride ${\rm UF}_4$ in a disproportionation reaction.

The degree of enriched $\alpha$  in isotope ${}^{235}{\rm U}$  (the degree of depleted $( \beta)$)
 or isotope ratio in the sample determined to the formula
\begin{equation}
\alpha  = \frac{{\nu {\rm (}{}^{{\rm 235}}{\rm U)}}}{{\nu {\rm (}{}^{{\rm 235}}{\rm U)} + \nu {\rm (}{}^{{\rm 238}}{\rm U)}}},\label{1}%
\end{equation}
where is $\nu {\rm (}{}^{{\rm 235}}{\rm U)}$  isotope substance amount  ${}^{235}{\rm U}$ ; is $\nu {\rm (}{}^{{\rm 238}}{\rm U)}$  isotope substance amount  ${}^{238}{\rm U}$.

Since, the averaged values degree of enriched or the isotope ratio: 1) the initial uranyl nitrate sample $\beta _{{\rm UO}_2^{2 + } }^{(0)}  = 0.00454$, 2) uranium tetrafluoride precipitate $\beta _{{\rm  UF}_4 }  = 0.00446$, and 3) regenerated uranyl nitrate $\alpha _{{\rm  UO}_2^{2 + } }  = 0.00464$. Therefore, in this work isotope separation ${{{}^{235}{\rm U}} \mathord{\left/
 {\vphantom {{{}^{235}{\rm  U}} {{}^{238}{\rm  U}}}} \right.
 \kern-\nulldelimiterspace} {{}^{238}{\rm U}}}$ enrichment factor equal:
 \begin{equation}
K^{(1)}  =  {\frac{{{{\alpha _{{\rm UO}_2^{2 + } } } {(1 - \alpha _{{\rm  UO}_2^{2 + } } )}}}}{{{{\beta _{{\rm UF}_4 } }  {(1 - \beta _{{\rm UF}_4 } )}}}}}   = 1.040, \label{2}
\end{equation}
 where  $K^{(1)}$ is the enrichment factor, and  $\alpha _{{\rm UO}_2^{2 + } }$ is the degree of enriched regenerated uranyl nitrate, $\beta _{{\rm UF}_4 }$ is the degree of depleted precipitate with ${\rm UF}_4$ .

Isotopic composition of regenerated uranyl nitrate and uranium tetrafluoride was measured by mass spectrometer. The enrichment factor of uranyl nitrate photolysed with MIE equal $1.04$, see formula~(\ref{2}).
\begin{figure*}
\includegraphics[height=8.5cm]{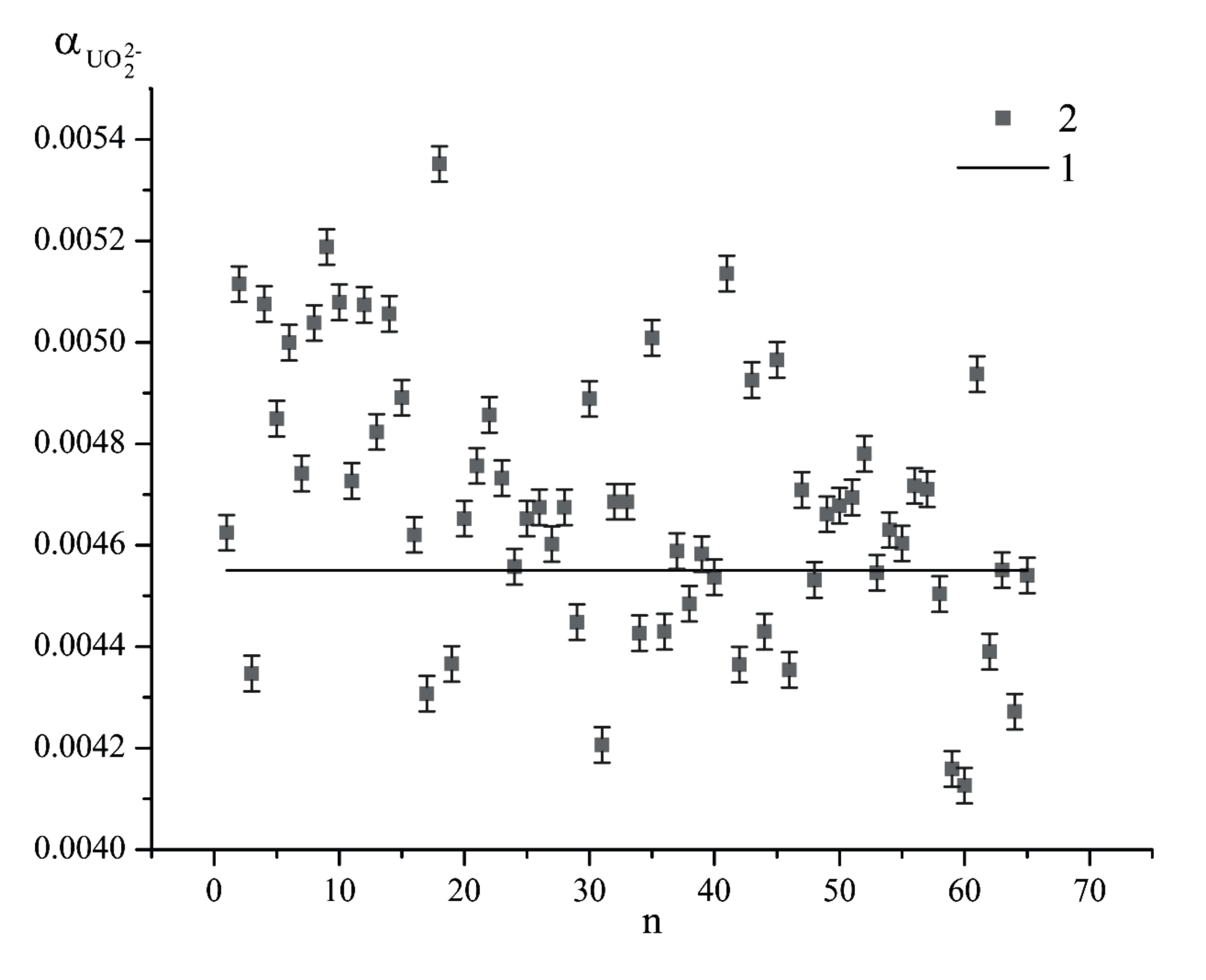}
\caption{Values of the degree of enriched $\alpha$   in isotope ${}^{235}{\rm U}$  from the 65 experiments (for each of value showed confidence interval with $\varepsilon  =  \pm {\rm 0}{\rm .00035}$):  degree of depleted of the initial uranyl nitrate sample $\beta  = {\rm 0}{\rm .00454}$  (1), and degrees of enriched (see formula~(\ref{1})) regenerated uranyl nitrate (\ref{2}).}\label{f2}
\end{figure*}

The experimental evidence of the degrees of enriched $\alpha$ with isotope ${}^{235}{\rm U}$ from the regenerated uranyl nitrate showed on Fig.~\ref{f2}. Here, we see 16 dots under the line corresponding to degree $\beta _{{\rm  UO}_2^{2 + } }^{(0)}  = 0.00454$, where uranyl nitrate depleted with isotope ${}^{235}{\rm U}$. Next, we see 8 dots allocated in a vicinity of the line corresponding to degree of the initial sample $\beta _{{\rm  UO}_2^{2 + } }^{(0)}  = 0.00454$. In the 41 experiments uranyl nitrate was enriched with isotope ${}^{235}{\rm U}$, as we see 41 dots above the line corresponding to degree $\beta _{{\rm  UO}_2^{2 + } }^{(0)}  = 0.00454$.

It follows from work~\citep{6Buchachenko} that enrichment factor  $K^{(1)}$  to have the value of $1.02$. In present work, the enrichment factor  $K^{(1)}$ to have the larger value $1.04$ (see formula~(\ref{2})) than value from paper~\citep{6Buchachenko}.

The enrichment factor related the separated methods of uranium isotopes are shown in papers~\citep{Sulaberidze,11Lysenko}. The Steenbeck’s centrifugal isotopes separations methods occupied a first place in accordance with efficiency. But, the MIE method of isotopes separation takes the second place. Unfortunately, for implement centrifugal isotopes separation method we necessary have high expenses for factory building, and for high technical complexity of construction.  Industrial method of isotopes separation by the gaseous diffusion is being implemented in the big industrial cluster includes plant and a power station. The building of such industrial cluster is a very expensive process. Therefore, the industrial method of isotopes separation by the gaseous diffusion is a difficult technical and economical problem.  An initial compound is uranium hexafluoride for the centrifugal isotopes separation methods and for the method of isotopes separation by the gaseous diffusion. In the period of terrorist danger, the plant with gaseous toxic uranium hexafluoride is a critical infrastructure object.

For the last 40 years, laser uranium separations have been investigated for the industrial processes~\citep{Sulaberidze}. Unfortunately, these investigations haven’t resulted in the commercial uranium enrichment technology.

Next, from papers~\citep{11Lysenko,12Lysenko} follows that the MIE method of uranium isotopes separation has a maximum enrichment factor except for centrifugal isotopes separation methods.

As of now, authors of the presented article are studying process of the uranium isotopes separation in the uranium oxides from natural samples. These processes of uranium isotopes separation have been induced MIE.

The isotope separation ${}^{235}{\rm U}$  from ${}^{238}{\rm U}$  in the photoinduced reaction uranyl nitrate with MIE in the water solutions demand further studying. Authors are studying kinetic properties ions, and stability of hydrate complexes with ions in water solutions.

The authors cordially thank Academician of the NAS of Ukraine V.\,G.\,Baryakhtar for his permanent attention and numerous discussions.

We are also grateful to our coauthors Corresponding Member of the NAS of Ukraine R.\,Ya.\,Belevtsev, and Professor V.\,V.\,Dolin.
We thank also Professor N.\,P.\,Malomuzh for the fruitful discussions of the results obtained.

\end{document}